\documentclass[twocolumn,prb,a4paper,amsmath,amssymb,floatfix,superscriptaddress,showpacs,10pt]{revtex4}
\usepackage{amssymb}
\usepackage{amsmath}
\usepackage{graphicx}
\usepackage{subfigure}
\usepackage{amsfonts}
\begin{document}

\title{Exchange interaction between magnetic adatoms on surfaces of noble metals}
\author{E.~Simon}
\affiliation{L\'or\'and E\"otv\"os University, Department of Physics, H-1518 Budapest POB 32, Hungary}
\author{B. \'Ujfalussy}
\affiliation{Research Institute for Solid State Physics and Optics of the Hungarian Academy of Sciences, Konkoly-Thege M. \'{u}t 29-33., H-1121 Budapest, Hungary}
\author{ B. Lazarovits}
\affiliation{Research Institute for Solid State Physics and Optics of the Hungarian Academy of Sciences, Konkoly-Thege M. \'{u}t 29-33., H-1121 Budapest, Hungary}
\author{A. Szilva}
\affiliation{Department of Theoretical Physics, Budapest University of Technology and Economics, Budafoki \'{u}t 8., H-1111 Budapest, Hungary}
\author{L.~Szunyogh}
\affiliation{Department of Theoretical Physics, Budapest University of Technology and Economics, Budafoki \'{u}t 8., H-1111 Budapest, Hungary}
\author{G.M.~Stocks}
\affiliation{Materials Science and Technology Division, Oak Ridge National Laboratory, Oak Ridge, Tennessee 37831, USA}

\begin{abstract}
We present first principles calculations of the exchange interactions between magnetic impurities deposited on (001), (110) and (111) surfaces of Cu and Au and analyze them, in particular, in the
asymptotic regime.
For the (110) and the (111) surfaces we demonstrate that the interaction shows an oscillatory behavior as a function of the distance, $R$, of the impurities and that the amplitude of the oscillations decays as 1/$R^2$. Furthermore, the frequency of the oscillations is closely related to the length of the Fermi vector of the surface states existing on these surfaces.
Due to the asymmetry of the the surface states' dispersion, the frequency of the oscillations
becomes also asymmetric on the (110) surfaces, while on the Au(111) surface two distinct
frequencies are found in the oscillations as a consequence of the Bychkov-Rashba splitting of the
surface states. Remarkably, no long range oscillations of the exchange interaction are observed
for the (001) surfaces where the surface states are unoccupied. When burying the impurities
beneath the surface layer, oscillations mediated by the bulk states become visible.
\end{abstract}

\maketitle

\section{Introduction}
The Rudermann-Kittel-Kasuya-Yoshida (RKKY) interaction in bulk materials is known for a long time, \cite{RKKY} and has become a textbook knowledge. In the past two decades, also the Oscillatory Exchange Coupling in magnetic multilayers has been extensively studied both in experiment \cite{oec1,oec2} and in theory \cite{bruno-chappert,Lathiotakis,edwards-umerski}. Recently, surface nanostructures containing only a few, or even just one atom were fabricated, and the exchange interaction between the individual atoms could be measured directly \cite{wiebe,zhou}. Various aspects of this interaction have already been discussed from first principles.\cite{stepa-cocu100a,stepa-cocu100b,stepa-cu111A,stepa-cu111B,wahl-cocu100,brovko-co2cu111}

In bulk, the magnitude of the RKKY interaction decays as $1/R^{3}$, where $R$ is the distance between the impurities, and the extremal vectors of the Fermi surface determine the frequency of the oscillatory interaction.
It is well known that the (111) surface of noble metals contains Shockley type surface states \cite{ss1,ss2,ss3,ss4}, which behaves as a two dimensional free electron gas (2DEG). It has been shown even earlier that the 2DEG mediates an RKKY type of interaction between
magnetic impurities which has the asymptotic form, $\sin(2k_FR)/R^2$,
where $k_F$ is the radius of the Fermi surface (circle).\cite{RZK-pr66,FischerKlein,LauKohn,Aristov-prb97,LitvDuga-prb98,hyldperss}

The study of realistic, more complicated (non-free-electron-like) Fermi surfaces or
the absence of a (partially) occupied
surface state call for more elaborate theoretical tools. Such methods account for the semi--infinite
host system beneath the surface, by calculating the surface Green's function as based on first principles.\cite{ecm-kkr,ecm-skkr} Previous studies of the exchange interaction at surfaces concentrated
merely on the (001) and (111) surfaces of the host material being mostly copper, \cite{stepa-cocu100a,stepa-cocu100b,stepa-cu111A,stepa-cu111B,wahl-cocu100,brovko-co2cu111} and a
comparative study of the vicinal surfaces of different host materials is still missing or incomplete, 
especially regarding the asymptotic regime.

In the present work, we perform calculations for the (100), (110) and (111) surfaces of Cu and Au in order to understand the role of various surface properties and their influence on the frequency and amplitude of the exchange interactions between Co atoms placed on these surfaces.
It is well-known that in case of  Au(111) the large spin--orbit interaction results in a splitting of the surface states' dispersion, called the Bychkov-Rashba splitting.\cite{BRS,ss1,ss3,ss4,henk} Here we demonstrate how
 the Bychkov-Rashba splitting manifests itself in the exchange interaction on Au(111) and also on Au(110).

\section{Method of Calculation}
The Screened Korringa-Kohn-Rostoker (SKKR) method\cite{skkr,skkr-book} combined with the embedding technique\cite{ecm-skkr} enables a precise treatment of a finite cluster of impurities embedded into a two-dimensional translational invariant semi-infinite host. Within multiple scattering theory, the electronic structure of a cluster of embedded atoms, $C$, is described by the corresponding scattering path operator (SPO) matrix, $\tau_{C}(\epsilon)$
($\epsilon$ being the energy), given by the following Dyson equation,~\cite{ecm-skkr}
\begin{equation}
\tau_{C}(\epsilon)=\tau_{h}(\epsilon)[1-(t_{h}^{-1}(\epsilon)-t_{C}%
^{-1}(\epsilon))\tau_{h}(\epsilon)]^{-1}\quad ,
\label{eq:tauc}
\end{equation}
where $t_{h}(\epsilon)$ and $\tau_{h}(\epsilon)$ denote the single-site scattering matrix and the
SPO matrix of the unperturbed host sites at the place of cluster $C$, respectively, while
$t_{C}(\epsilon)$ stands for the single-site scattering matrix of the embedded atoms. 
 Self-consistent calculations within the local density approximation (LDA) of the density functional theory (DFT)
 can then be easily performed.\cite{skkr-book}  
 In the present calculations no attempt was made to include surface relaxations: the
  cluster and the host sites refer to the positions of an ideal FCC lattice with the experimental lattice constant of Cu ($a_{\rm fcc}=3.6147$ \AA ) and Au ($a_{\rm fcc}=4.0648$ \AA ).
First, we performed self-consistent (relativistic) calculations for the (001), (110)
and the (111) surfaces of Cu and Au, then for Co impurities placed on the top of these
surfaces. The details of the scf calculations are described in Ref.~\onlinecite{ecm-skkr}.

Once the self-consistent potential and exchange field for a single impurity has been obtained, we used
the magnetic force theorem\cite{Jansen99} to calculate the interaction between two magnetic impurities.
As in this procedure the relaxation of the electronic structure due to the proximity
of the impurities is neglected, the exchange interaction between two
impurities should be calculated from the grand canonical potential, $\Omega=E_{\rm band} -
\epsilon_F N$, with $E_{\rm band}$, $\epsilon_F$ and $N$ being the band-energy, the Fermi level
 and the number of valence electrons, respectively.

For uniaxial systems, such as considered in this paper, the dependence of the grand potential
on the orientations of the magnetic moments,  $\vec{e}_1$ and $\vec{e}_2$, can be
written up to second order as\cite{julie88,cr3}
\begin{equation}
\Omega(\vec{e}_1,\vec{e}_2;\vec{R})= \Omega_0(\vec{R})
+ K \left( e_1^z \right)^2 + K \left( e_2^z \right)^2
+  \frac{1}{2} \, \vec{e}_1 \,{\bf J} (\vec{R}) \, \vec{e}_2 \; ,
\label{eq:om12}
\end{equation}
where $K$ is the uniaxial anisotropy constant of a single-impurity and
${\bf J} (\vec{R})$ is a $3 \times 3$ matrix comprising the isotropic,
symmetric anisotropic and the asymmetric exchange (Dzyaloshinskii-Moriya) interactions between
two impurities placed at a relative position, $\vec{R}$.

As usual, we define the exchange coupling energy between two impurities as
\begin{equation}
J(\vec{e};\vec{R}) = \Omega(\vec{e},\vec{e};\vec{R})-\Omega( \vec{e},-\vec{e};\vec{R})
\; ,
\label{eq:Je}
\end{equation}
a quantity that still depends on the orientation, $\vec{e}$.
From Eq.~(\ref{eq:om12}) we, however, see that
\begin{equation}
J(\vec{e};\vec{R}) = J(\vec{R}) + \vec{e}\, {\bf J}^{\rm sym} (\vec{R}) \,\vec{e} \; ,
\label{eq:Je2}
\end{equation}
where
\begin{equation}
J(\vec{R}) = Tr{\bf J} (\vec{R})/3 \; ,
\label{eq:Jiso}
\end{equation}
is the isotropic exchange coupling parameter and
\begin{equation}
{\bf J}^{\rm sym} (\vec{R}) = \frac{1}{2} \left( {\bf J}(\vec{R}) + Tr{\bf J}(\vec{R}) \right)
- J(\vec{R}) {\bf I} \; ,
\label{eq:Jsym}
\end{equation}
is the traceless symmetric part of ${\bf J} (\vec{R})$ with the unit matrix ${\bf I}$.
It should be noted that the so-called
symmetric exchange (two-site anisotropy), ${\bf J}^{\rm sym} (\vec{R})$, arises purely from
relativistic effects and it
scales with the squared spin-orbit coupling strength. Since our calculations indicate that
it is by several orders smaller in magnitude than the isotropic exchange coupling, $J(\vec{R})$,
in this study we disregard the orientational dependence of $J(\vec{e};\vec{R})$ and
present calculation of Eq.~(\ref{eq:Je}) at the specific orientation, $\vec{e}=\hat{z}$.
In order to carefully trace the long-range oscillations of $J(\vec{R})$, when calculating
the host $\tau_{h}$ matrices, see Eq.~(\ref{eq:tauc}), we used 30 000 $k_{||}$ points 
in the irreducible wedge of the surface Brillouin zone, while 
10 000 $k_{||}$ points were sufficient to use in the absence of occupied surface states.

In Table I, calculated exchange interactions are shown 
between impurities at selected distances for the case of a Cu(111) surface.
Note that negative/positive values indicate ferromagnetic/antiferromagnetic
coupling between the spins.
Two approaches are compared:  {\em (i)} using single impurity potentials as described above and
{\em (ii)} taking into account the proximity of the two impurities in terms of self-consistent calculations.
We denote the respective results for the exchange interactions by  $J_{1i}(\vec{R})$ and $J_{2i}(\vec{R})$.  
As expected, for impurities at large distances ($R=10\, a$ and $20\, a$, where $a$ is the nearest
neighbor distance) the two methods are in reasonable agreement. 
For the nearest neighbor pair,
however, the exchange energy obtained from 
self-consistent two-impurity potentials are roughly twice as big as the ones calculated from
the single-impurity approach. 
\begin{table}[ht]
\begin{center}
\begin{tabular}{ccc}
$R/a$ & $J_{1i}$ & $J_{2i}$ \\
\hline
\hline
1 & -197 meV  & -106 meV  \\
10 & -24.5 $\mu$eV & -24.3 $\mu$eV\\
20 & 1.29 $\mu$eV & 1.31 $\mu$eV \\
\end{tabular}
\caption[smallcaption]{Exchange interactions between Co impurities at selected distances, $R$ (in
units of the nearest neighbor distance, $a$), on a Cu(111) surface as calculated by two different approaches (see text).}
\label{table1}
\end{center}
\end{table}

\section{Results and discussion}
\subsection{Nearest neighbor interactions}
Although the main purpose of the current paper is to investigate the asymptotic behavior of the magnetic interaction between two surface impurities, we would like to briefly comment on the nearest neighbor interactions as well. For two impurities that are close enough to have sufficient overlap of their wave functions, it is the direct exchange which dominates the interaction. It gives a strong, but short range coupling which decreases rapidly.  We compare the exchange energy for nearest neighbor impurities in bulk and on surface as calculated from the single impurity approach (see above).  As we have seen,
for the nearest neighbor pairs this is a crude approach, but it provides a qualitative
estimate for the desired comparison.

Our results for Cu and Au hosts are presented in Table \ref{table2}.  In both cases
the coupling of the two Co spins are ferromagnetic in the bulk. This interaction is
largely enhanced at the  (100) and (111) surfaces. In general, this enhancement
can be correlated  with the decreased number of host atoms in nearest neighbor
position below the Co atoms, 4 for (100) and 3 for (111), correspondingly, with a decreased
hybridization between the Co and host atoms. 
 Though very small in magnitude, on top of the (110) surface the nearest neighbor 
 exchange interactions become antiferromagnetic. This observation clearly demonstrates that
 the surface electronic structure can dramatically change the interactions between
 adatoms as compared to the bulk.
\begin{table}[ht]
\begin{center}
\begin{tabular}{cccc}
Host & Surface & $J_{\rm surf} (meV)$ & $J_{\rm bulk} (meV)$ \\
\hline
\hline
 & 100 & -198 &  \\
Cu & 110 & 10.5 & -48.9 \\
 & 111 & -197 &  \\
\hline
 & 100 & -109 &  \\
Au & 110 & 7.85 & -38.5 \\
 & 111 & -135 &  \\
\end{tabular}
\caption[]{(color online) Calculated nearest neighbor exchange interactions in bulk and on different surfaces of Cu and Au. }
\label{table2}
\end{center}
\end{table}

\subsection{Asymptotic behavior}

\begin{figure}[htb]
\begin{center}
\includegraphics[width=8.5cm,bb=92 42 318 296,clip]{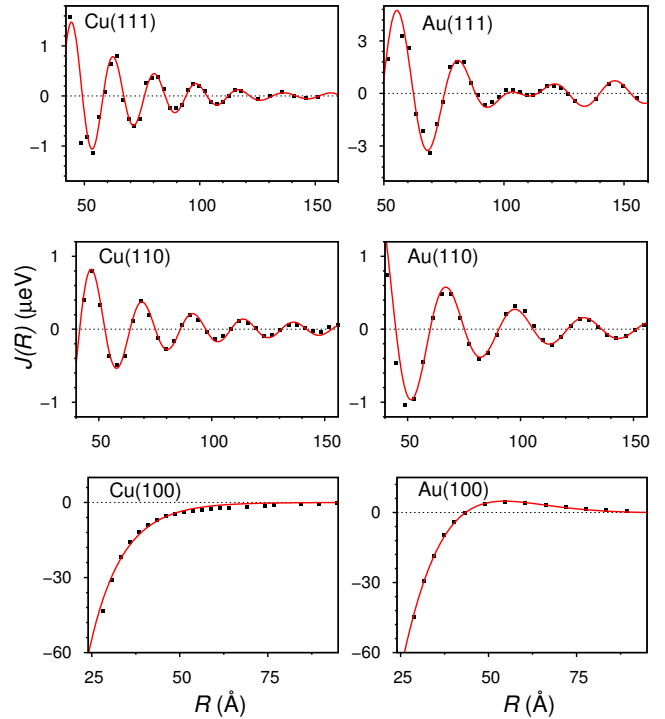}
\end{center}
\caption{(color online) Calculated exchange interactions as a function of the distance
between Co adatoms on top of the vicinal surfaces of  Cu and Au  along the directions 
specified in the text. Symbols refer to the calculated data, solid lines to the fitted curves.}
\label{exch-all}
\end{figure}

We calculated the exchange interactions, $J(R)$, between two Co adatoms deposited on top
of the (111), (110) and (100) surfaces of Cu and Au surfaces for distances up to
 $R \simeq 100-150$ \AA, to be considered safely as the asymptotic region.
 The calculated results are shown in Fig.\ref{exch-all} as a function of the distance between the adatoms. 
 In case of the (100) and the (111) surfaces, the two Co atoms were placed along the (110), i.e.,  the 
 nearest neighbor direction, while in case of the (110) surface along the (001), i.e., the next nearest 
 neighbor direction.
 
For the (111) surface of Cu, see left-top panel of Fig. \ref{exch-all}, $J(R)$  shows clear oscillations with a period of $L=18.5$ \AA, corresponding to a Fermi wavelength of
the surface states, $k_F=\pi/L=0.17$ 1/\AA.
This period is very close to the value obtained from STM measurements and also from
first-principles calculations
by Stepanyuk {\em et al.}, $L=15$ \AA \ or $k_F=0.22$ 1/\AA .\cite{stepa-cu111A,stepa-cu111B}
The difference between the theoretical results can mostly be attributed to different angular
momentum cut-offs and to different surface potentials used in the calculations.
In agreement with theoretical models\cite{RZK-pr66,FischerKlein,LauKohn,Aristov-prb97,LitvDuga-prb98,hyldperss}, our numerical fit also confirmed
that the amplitude of the oscillations decays as 1/$R^2$.

The asymptotic curve of $J(R)$ displays a more complicated behavior for Au(111), see
right-top panel of  Fig. \ref{exch-all}, since this curve could be fitted as the
sum of two oscillations with the Fermi wavelengths,
$k_F^1=0.104$ 1/\AA \ and $k_F^2=0.142$ 1/\AA. The appearance of the two oscillation
periods is due to the famous Bychkov-Rashba splitting\cite{BRS} of the Au(111)
surface states experiencing the strong spin-orbit interaction of Au. This splitting of the
surface states gives rise to two distinct spherical Fermi-cuts, thus, to two
distinct asymptotic oscillations for the exchange interaction between magnetic adatoms.
It should be noted, however, that, mainly because of the imprecise treatment of the
surface potential within the ASA, our values for the Fermi wavelengths are typically smaller,
 while their difference, $\Delta k_F=0.038$ 1/\AA, is larger as reported in the literature. \cite{ss1,ss2,ss3}

As studied by Petersen {\em et al.} in terms of Fourier transform STM,\cite{petersen-cu110}
there exist partially occupied Shockley states on the Cu(110) surface that are located
at the boundary  ($\overline{\rm Y}$ point) of the two dimensional Brillouin zone. 
These surface states naturally
give rise to oscillatory exchange interactions between magnetic adatoms as demonstrated
in the middle panels of Fig. \ref{exch-all}, both for Cu(110) and Au(110).

\begin{figure}[htb]
\begin{center}
\includegraphics[width=8.5cm,bb=55 49 215 160,clip]{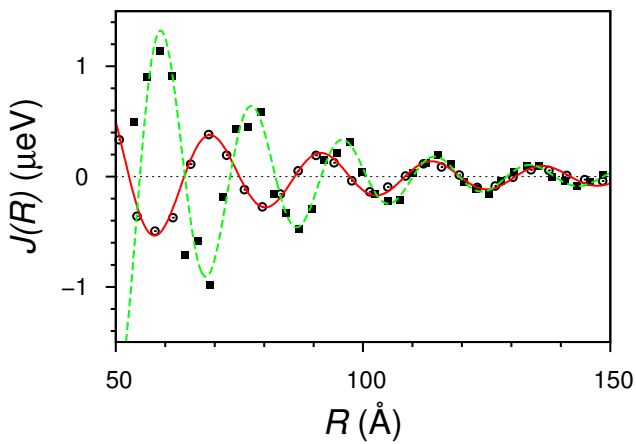}
\caption{(color online) Calculated exchange interaction on the Cu(110) surface along two different directions:
 (001) (circle) and (1$\bar{1}$0) (filled square). The solid and dashed lines
 correspond to the fitted curves.}
\label{dirs}
\end{center}
\end{figure}

Noticeably, the shape of the cuts of these surface states is elliptic due to the $C_{2v}$ symmetry of the surface.  Consequently, the period of the oscillations depends on the direction along which the two adatoms are pulled apart.\cite{esimon1} 
Fig. \ref{dirs} presents the exchange interactions
in case of Cu(110) along the (001) and the (1$\bar{1}$0) directions.
The Fermi wavelengths obtained from fitting these curves, $0.141$ 1/\AA \ and $0.172$ 1/\AA  \
along the (001) and the (1$\bar{1}$0) directions, agree indeed very well with the values derived
directly from the dispersion relation of the surface states, $0.138$ 1/\AA  \ and $0.164$
1/\AA \, respectively.

In case of Au(110), we observed two oscillations in $J(R)$ along the (001) direction, corresponding to the Fermi wavelengths, $0.098$ 1/\AA \ and $0.118$ 1/\AA . 
Similar to Au(111), this can again be explained by the fact that the Au(110)
surface states  experience an anisotropic Bychkov-Rashba splitting.\cite{esimon2} 
Note, however, that the amplitude of the oscillation with the longer period ($k_F =  0.098$ 1/\AA )
turned out to be about 16 times larger than the one for the shorter period   ($k_F =  0.118$ 1/\AA ).
Most likely, an even larger imbalance of the amplitudes applies to the asymptotic 
exchange interactions along the direction (1$\bar{1}$0), 
since in this case it was not possible to resolve numerically the 
two different frequencies in the oscillations: our fit confirmed only the short period with $k_F=0.151$ 1/\AA .

In obvious contrast to all previous cases, in case of the (100) surfaces (lower panels of
Fig. \ref{exch-all}) the exchange interactions show
no oscillations in the asymptotic regime. For Cu(100), $J(R)$ is well described by
an exponential decay. For Au(100), it also rapidly decreases with $R$, though,
$J(R)$ changes sign at $R \simeq 30$\AA. Therefore, we may conclude that, lacking
(occupied) free-electron type surface states, there is no long-range RKKY type exchange interaction between the magnetic atoms on the (100) surfaces. At first glance, this statement contradicts to the observation of Stepanyuk {\em et al.}\cite{stepa-cocu100b} who established an
oscillatory coupling even in case of Cu(100). Their calculations were, however,
restricted to distances, $R < 10$ \AA, that can not be regarded as the asymptotic regime.

\begin{figure}[htb]
\begin{center}
\includegraphics[width=8.5cm,bb=89 44 320 212,clip]{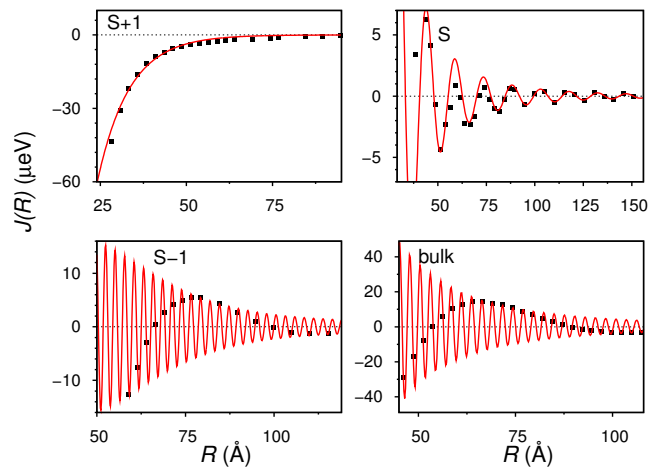}
\caption{(color online) Calculated exchange interactions between two Co impurities placed at different
vertical positions with respect to a Cu(100) surface: $S+1$ on top, $S$ in the surface layer,
$S-1$ in the subsurface layer,  as well as in the bulk. }
\label{surface-lays}
\end{center}
\end{figure}

However, once the Co adatoms are placed in the surface layer, an oscillatory exchange interaction appears, although  with a different frequency than in the bulk, see right-top panel of Fig. \ref{surface-lays} for the case of Cu(100).  When placing the impurities even deeper beneath the surface, 
the frequency of the bulk RKKY interaction is quickly recovered.
It is well-known that the Cu Fermi surface has four extremal vectors along the
(110) direction.\cite{Lathiotakis} Among them, the largest one, related
to the (110) diameter of the 'dog-bone'--shaped Fermi surface,  causes the rapid
oscillations in the bulk as seen on the bottom-right panel of Fig. \ref{surface-lays}. Interestingly, our numerical analysis shows that the frequency
of the oscillations for adatoms in the surface ($S$) layer correlates with the much shorter
extremal vector of the Fermi--'neck'.

\begin{figure}[htb]
\begin{center}
\includegraphics[width=8.0cm,bb=54 50 207 160,clip]{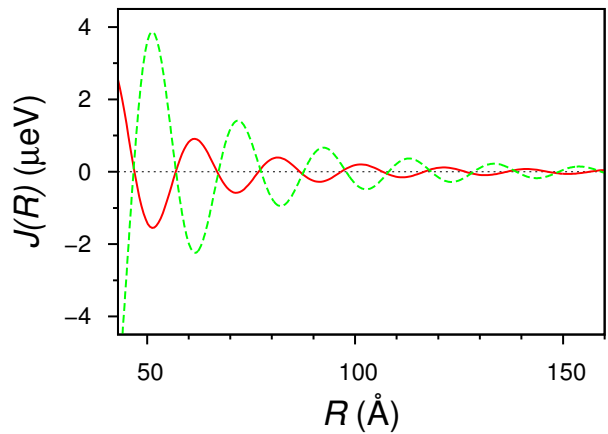}
\caption{Fitting curve of the calculated exchange interactions between two Fe (dashed line) and Co (solid line) 
adatoms placed on top of a Cu(110) surface along the (1$\bar{1} \bar{1}$) direction. }
\label{imps}
\end{center}
\end{figure}

So far, we presented numerical results for the exchange interaction between two Co impurities.
In Fig. \ref{imps} the exchange interactions between two Co adatoms are compared
with that between two Fe adatoms placed on top of Cu(110) along the (1$\bar{1} \bar{1}$) direction.
From this figure, it can clearly be inferred that the frequency of the asymptotic oscillations is the same for both cases: the type of the adatoms becomes manifest in the amplitude and in the phase of the oscillations. The period of $L=20.4$ \AA \ corresponds to
a Fermi wavelength of $k_F=0.154$ 1/\AA \ which, reassuringly, coincides well 
with the length of the spanning vector of the elliptical Fermi surface cut along the corresponding direction.

\section{Conclusion}
We presented first principles calculations of the exchange interactions between magnetic impurities deposited on the vicinal surfaces of Cu and Au. 
In full agreement with previous theoretical studies, for the (110) and the (111) surfaces we demonstrated that the interaction in the asymptotic regime is oscillatory: 
the amplitude decays as 1/$R^2$, where $R$ is the distance of the adatoms, and 
the frequency of the oscillations coincides with the length of the Fermi vector 
of the Shockley type surface states existing on these surfaces.
In case of the (110) surfaces, the surface states around the $\bar{\rm Y}$ point of the surface
Brillouin zone have an elliptic paraboloid
dispersion relation, due to the $C_{2v}$ point-group symmetry. This, in turn, resulted in an anisotropic 
periodicity of the exchange interaction when varying the direction between the adatoms. 
Moreover, for the Au(111) surface two distinct
frequencies are found in the oscillations at any direction as a consequence of 
the Bychkov-Rashba splitting of the surface states. 

Our most remarkable observation is the
lack of long range oscillations in the asymptotic exchange interaction 
for the (001) surfaces. This finding can be correlated with the fact that there are 
no partially occupied surface states in this case. We should note that the bulk Fermi-surface related oscillations decaying as 1/$R^5$ predicted by Lau and Kohn~\cite{LauKohn} could not be
numerically resolved in our calculated data.
Burying, however, the impurities
beneath the surface layer, oscillations mediated by the bulk states become apparent.

\acknowledgments
Financial support was provided by the Hungarian Research Foundation (contract no. OTKA K68312 and K77771)
and by the New Hungary Development Plan (Project ID: T\'AMOP-4.2.1/B-09/1/KMR-2010-0002).

\end{document}